\documentclass{aa}
\usepackage{graphicx}
\begin{document}
\title{\bf Genetic Connections of Anomalous X-ray Pulsars and Soft Gamma 
Repeaters with Supernova Remnants}
\titlerunning{Genetic Connections of AXPs \& SGRs with SNRs}

\author{A. Ankay\inst{1}
\and
 O. H. Guseinov\inst{2}
\and
 M. A. Alpar\inst{3}
\and
 S. O. Tagieva\inst{4}
}
\authorrunning{Ankay et al.}

\offprints{A. Ankay}

\institute{Department of Physics, Middle East Technical University, 
Ankara 06531, Turkey \\
\email{askin@astroa.physics.metu.edu.tr}
\and
Department of Physics, Akdeniz University, Antalya, Turkey \\
\email{huseyin@pascal.sci.akdeniz.edu.tr}
\and
Faculty of Engineering and Natural Sciences,
Sabanc\i\ University, Orhanl\i\ - Tuzla, \.{I}stanbul 81474, Turkey \\
\email{alpar@sabanciuniv.edu}
\and
Academy of Science, Physics Institute, Baku 370143, Azerbaijan Republic \\
\email{physic@lan.ab.az}}

\date{Received; accepted}

\abstract
{Genetic connections between anomalous X-ray pulsars (AXPs) or soft gamma
repeaters (SGRs) and supernova remnants (SNRs) can hold important clues to
the nature of these objects through the properties of the associated SNRs. 
We examine various criteria jointly to assess the reliability of possible
connections. 3 AXPs and 2 SGRs are most probably genetically connected
with SNRs. Distances and ages are discussed and their likely ranges are
indicated. 
}

\maketitle

\section{Introduction} 

In this work, we examine the connections of
anomalous X-ray pulsars (AXPs) (see van Kerkwijk 2000 and Mereghetti 
2001a, 2001b for reviews)
and soft gamma repeaters (SGRs) (see Hurley 2000 for
a review) with supernova remnants (SNRs).

These two classes of objects have attracted much attention. The SGRs
exhibit repeated bursts of soft gamma rays. Their quiescent X-ray
properties are similar to those of AXPs. Indeed, AXPs and SGRs together 
constitute the set of all
neutron stars with typical accretion luminosities, observed rotation 
periods but without evidence for binary companions. Both classes of objects 
are similar in their periods (P $\sim$ 10 s), period derivatives, spectra, 
and luminosities. 

There are currently two alternative models for these objects. The magnetar
model posits that these sources are isolated rotating neutron stars with 
surface dipole magnetic fields of 10$^{14}$-10$^{15}$ G, extracting their
luminosity from the decay of the magnetic field. The magnetar picture also
provides a model for the soft gamma-ray bursts. The spindown of the object is
believed to obey a modified dipole spindown equation,
$\dot{\Omega}$ = --k$\Omega$$^n$ where n = 3 for pure dipole radiation, and
n can be greater than 3, possibly as high as n = 5 when the decay of the
magnetic field is driven by spindown (Ruderman et al., 1998). In magnetar 
models the age of the neutron star can be estimated as a characteristic time 
$\tau$ = $P/(n-1)\dot{P}$, assuming that the initial rotation period was 
much shorter than the current rotation period.

The alternative model is that the X-ray luminosity is due to accretion
from a fall-back disk that is left over from the supernova that formed the
neutron star. This model does not include a dynamical explanation for the 
gamma-ray
bursts of the SGRs. It seeks to classify all kinds of
young neutron stars as propellers and accretors from a fall-back disk. The
similarities and differences of the different classes, including the radio
pulsars, the dim thermal neutron stars, and the "radio quiet" neutron stars 
like the source in
Cas A in addition to the AXPs and perhaps SGRs, are supposed to be due to
the presence or absence of a fall-back disk and different disk initial
conditions. In this model AXPs and SGRs are supposed to be in an
asymptotic regime near rotational equilibrium, and there is no way to
estimate their ages from the observed period, P and period derivative,
\.{P}. Therefore, for discussing the possible association of AXPs and
SGRs with SNRs, age information for the SNR can be used to make a 
comparison only if it is assumed that the AXP or SGR is a magnetar. 

Finding genetic relations between AXPs/SGRs and SNRs is important
for several reasons. We have much more information
about SNRs than we do for AXPs/SGRs. If we find a genetic
connection between an AXP/SGR and a SNR we can use
the SNR's data for the AXP/SGR. In particular
the age of the SNR can be used to provide an age estimate for the
AXP/SGR within error limits. It is also important to find
distances of AXPs/SGRs. By using the distance value of the associated
SNR as the distance of the AXP/SGR we can calculate the luminosity of such
objects.
Further, it is important to compare the characteristics of the
SNRs possibly connected with AXPs/SGRs with the characteristics of other
SNRs associated with radio pulsars to see if there are
significant differences between them.                                          

Some of the young radio pulsars are not associated with 
SNRs; Kaspi (2000) gives a list of the youngest 17 pulsars with 
"characteristic ages" $\tau$ = P/2\.{P} $\le$ 2.44 $\times$ 10$^4$ yrs. For 
7 of these pulsars 
genetic relations with SNRs were not found. It can therefore be inferred
that some SNRs have very short lifetimes. 
Therefore, failing to find
SNRs around AXPs and SGRs does not necessarily mean
that AXPs and SGRs have to be much older than 10$^4$-10$^5$ yrs, the
lifetime of SNRs.

There are several criteria to be checked to see if there is a genetic
connection between an X-ray pulsar and a SNR (see Kaspi 2000 and 
Allakhverdiev et al. 1997a for pulsar - SNR connections).
One must try to use all of these criteria to check if the connection is 
real or not, taking into account the reliability and applicability of each 
of the criteria. 
Some important points to be checked for an association to be real are: 
1) Position of the neutron star with respect to the geometrical center of 
the SNR. 
2) Evidence for interactions between the neutron star and the ambient medium.
3) Comparing neutral hydrogen column density (N$_{\hbox{\footnotesize 
{HI}}}$) values of the neutron star and the SNR within error limits to 
check if they can be at the same distance.
4) Comparing the estimated space velocities of AXPs/SGRs with the 
space velocity distribution of radio pulsars. The velocity estimates in 
our Tables refer to the projected space velocity on the sky and, on the 
average, space velocity is (3/2)$^{1/2}$ times the projected space velocity.
5) Comparing the age of the SNR and the characteristic time, 
$\tau$ = $P/(n-1)\dot{P}$ of the AXP/SGR is relevant for assessing the 
possibility that these sources are magnetars. The characteristic time 
taken as $\tau$ is close to the real age only if the neutron star is an 
isolated rotating dipole and if its period is much greater than its 
initial period. If AXPs and SGRs are magnetars then, due to magnetic field 
decay and/or interaction with the surrounding matter, a characteristic 
age $\tau$ = $P/(n-1)\dot{P}$, n $\not=$ 3, may give a close estimate of the 
real age. The comparison of the neutron star age with the SNR age is 
model dependent but possible for a range of magnetar/field-decay models.  
If AXPs and SGRs are accreting sources, then $\tau$ = $P/(n-1)\dot{P}$ is 
not related with the real age. 

Among these criteria the strongest one is the position of the AXP/SGR with
respect to the SNR.
Because AXPs and SGRs are rare objects, chance projection is 
not very probable. The data related with the other criteria 
are usually insufficient and not very reliable. As the 
observational data improve, the criteria other than
the position will become more usable. Our strategy is to take all
criteria into account while examining the connections,
keeping in mind the uncertainties of each criterion. The dynamical
characteristics of the shells of the SNRs possibly connected with 
the AXPs/SGRs are, in general, similar to the characteristics of the 
radio pulsar associated SNR shells. We therefore assume, for our
discussion of space velocities, that AXPs/SGRs have roughly the same
kinematical characteristics as the radio pulsars. 

In a companion paper we attempt to improve the distance values of the SNRs 
given in the Galactic SNRs catalog (Green, 2000) by examining the distance 
values of the SNRs given in the literature, their neutral hydrogen column
densities (N$_{\hbox{\footnotesize {HI}}}$), explosion energies (E) of the
supernovae that formed them, and densities of the ambient medium 
(Ankay et al., 2001). The distance and the diameter values presented without 
giving references in Tables 2 and 4 were taken from this work.

In Section 2, we discuss possible connections of SGRs and AXPs to
SNRs. For AXP/SNR associations our conclusions weighing all available
criteria agree with the conclusions of Gaensler et al. (2001) on the basis
of positional coincidence alone. For SGR/SNR associations, we conclude
that several connections previously excluded by Gaensler et al. (2001) are
actually likely. Marsden et al. (2001) also examined possible connections
for the group of AXPs and SGRs we investigate here. In addition, they 
study the sources AXP 0720-3125 and SGR 1801-23 which are not confirmed as 
AXP and SGR, respectively. These two sources are not considered here.
They conclude almost all of the AXPs/SGRs, and the unconfirmed AXP/SGR 
candidates they study are associated with SNRs, leading in some cases to 
very high neutron star space velocities with realistic distances. 
These associations were adopted without comparative study of different 
criteria. We conclude that only 3 AXP-SNR associations and 2 SGR-SNR 
associations can be adopted with some confidence. We present the 
conclusions in Section 3.

\begin{table*}
\centering
\caption[]{The Data of Soft-gamma Repeaters}
\label{table1}
$$
\begin{array}{p{0.2\linewidth}ccccc}
\hline
\noalign{\smallskip}
Names \& & $P$ & $\.{P}$ 10^{-11} & \tau 10^3 & $d$ & 
$N$_{\hbox{\footnotesize {HI}}} 10^{21} \\ 
Remarks & ($s$) & ($ss$^{-1}) & ($yr$) & ($kpc$) & ($cm$^{-2}) \\
& & & & & (0.1-2.4 $keV$) \\ 
            \noalign{\smallskip}
            \hline
            \noalign{\smallskip}            
SGR 0526-66 & 8.1 & & & 50-55 & 5-10 \\
SNR N49 & [6] & & & & [7] \\
Connection & & & & & \\ \hline
SGR 1806-20 & 7.47 & 8.3 & 1.4 & 14.5 & 60 \\
(AX 1805.7-2025) & [1] & [1] & & [15] & [1, 10, 12] \\
SNR G10.0-0.3 & & 8.1-11.7 & & & \\
Jet & & [9] & & & \\
$[14, 13]$ & & & & & \\
Connection & & & & & \\ \hline
SGR 1900+14 & 5.16 & 11 & 0.74-1.3 & 5 & 21.6 \\
SNR G42.8+0.6 & [2, 3, 8, 13] & [2] & 0.71 & [3] & [3] \\
& & 5.82 & [3] & & \sim30-50 \\
& & [4] & & & [2] \\
& & 6.0 & & & \\
& & [13, 10] & & & \\
& & 6.13 & & & \\
& & [8] & & & \\ \hline
SGR 1627-41 & 6.41? & & & 11 & 77 \\
SNR G337.0-0.1 & [4, 5] & & & [11] & [4] \\
(CTB 33) & & & & & \\ \hline
            \noalign{\smallskip}
         \end{array}
     $$
\begin{list}{}{}  
\item[$^{\mathrm{}}$] 
[1] Kouveliotou et al., 1998; [2] Kouveliotou et al.,
1999; [3] Hurley et al., 1999b;
[4] Woods et al., 1999a; [5] Dieters et al., 1998;
[6]Mazets et al., 1979; [7] Marsden
et al., 1996; [8] Woods et al., 1999b; [9] Woods et
al., 2000; [10] Sonobe et al., 1994;
[11] Corbel et al., 1999; [12] Murakami et al., 1994;
[13] Marsden et al., 1999; [14] Kulkarni et al., 1994; [15] Green, 2000
\end{list}
   \end{table*}

\section{Connections of AXPs and SGRs with SNRs}
Below, we examine each pair of AXP/SGR-SNR using the data of AXPs/SGRs 
and SNRs given in Tables 1,3 and 2,4, respectively. In the Tables, 
$\beta$ $\equiv$ $\Delta$$\theta$/$\theta$ is the ratio between the offset, 
$\Delta$$\theta$,
of the neutron star's position from the geometrical center of the SNR and 
the angular radius, $\theta$, of the SNR. By this definition, 
$\Delta$$\theta$/$\theta$ corresponds to V$_{\hbox{\footnotesize 
{NS}}}$/V$_{\hbox{\footnotesize {exp}}}$, where V$_{\hbox{\footnotesize 
{NS}}}$ denotes the projected space velocity of the neutron star and 
V$_{\hbox{\footnotesize {exp}}}$ denotes the expansion velocity of the 
SNR.

\begin{table*}
\centering
\caption[]{The Data of Supernova Remnants in the Directions of SGRs}
\label{table2}
$$
\begin{array}{p{0.1\linewidth}ccccccccccccc}   
\hline
\noalign{\smallskip}
Names \& & d & t & E_{\hbox{\footnotesize {kin}}} & kT & 
Type & \alpha & \Delta\theta/\theta & D & 
N_{\hbox{\footnotesize {HI}}} & L_{\hbox{\footnotesize
x}} & Density & V_{\hbox{\footnotesize {NS}}} \\  
Remarks & (kpc) & (kyr) & (10^{50} erg) & (keV) & & & (or \beta) & 
(pc) 
& (10^{21} cm^{-2}) & (10^{33} erg/s) & (\sharp/cm^3) &
(km/s) \\ \hline 
            \noalign{\smallskip}
            \hline
            \noalign{\smallskip}
N49 & 50-55 & 5.5 & \sim15 & 0.6 & S? & & 0.6-1 & 16 &
21 & 63^{\hbox{\footnotesize a}} & n_{\hbox{\footnotesize 
o}}=30^{\hbox{\footnotesize c}} & 1200^{\hbox{\footnotesize e}} \\ 
(in LMC) & & [8,11,15] & [15] & [15] & & & [8] & [8] &
[15] & [15] & [14,8] & [11] \\
& & & & & & & 0.8 & & & 2100^{\hbox{\footnotesize b}} & 
n_{\hbox{\footnotesize o}}=2.6^{\hbox{\footnotesize c}} & 2900 \\
& & & & & & & [11] & & & [13] & [15] & [10] \\
& & & & & & & & & & & 20-940^{\hbox{\footnotesize d}} & \\
& & & & & & & & & & & [13] & \\ \hline
G10.0-0.3 & 12 & 10 & & & ? & 0.8 & 0.0-0.5 & 24 & & & Dense & 800 \\
& 11 & [11] & & & F & [18] & 0.5 & & & & Maser & [11] \\
& [9] & & & & [6] & 0.6 & [11] & & & & [1] & \\
& 14.5 & & & & & [6] & & & & & & \\
& [17] & & & & & & & & & & & \\ \hline
G42.8+0.6 & 6 & 10 & & & S & 0.5? & 1.2-1.4 & 42 & & 
^{\hbox{\footnotesize f}} & Not & \sim1800 \\
& 5 & [6,10,11] & & & & [18] & [10] & & & [7] & Dense & [10] \\
& [7] & & & & & & 1.4 & & & & & 2000 \\
& & & & & & & [11] & & & & & [11] \\
& & & & & & & & & & & & 480-5500 \\
& & & & & & & & & & & & [5] \\ \hline
G337.0-0.1 & 12 & \sim5 & & & S & 0.6? & 2-2.3 & 6 & & & Dense? &
800 \\
(CTB 33) & 11 & [9] & & & & [18] & [9] & 4.8 & & & Masers & [10] \\
& [9,4] & & & & & & 1.6 & [9] & & & [12,1,9,2] & 1000 \\
& & & & & & & [3] & 5.1 & & & & [11,9] \\
& & & & & & & 1.7 & [4] & & & & 200-2000 \\
& & & & & & & [16] & & & & & [16] \\ \hline

            \noalign{\smallskip}
         \end{array}  
     $$

\begin{list}{}{}
\item[$^{\mathrm{}}$]

[1] Frail et al., 1996; [2] Brogan et al., 2000;
[3] Smith et al., 1999; [4] Sarma et al., 1997; [5] Hurley et al.,
1999b; [6] Vasisht et al., 1995; [7] Vasisht et al., 1994;
[8] Vancura et al., 1992; [9] Corbel et
al., 1999; [10] Gaensler, 2000; [11] Marsden et al., 1999;
[12] Koralesky et al., 1998b; [13] Banas et al., 1997;
[14] Blair et al.,
2000; [15] Hughes et al.,
1998; [16] Hurley et al., 1999a; [17] Green, 2000 

\end{list}

\begin{list}{}{}  
\item[$^{\mathrm{a}}$] for 0.5-5 keV
\item[$^{\mathrm{b}}$] optical
\item[$^{\mathrm{c}}$] preshock
\item[$^{\mathrm{d}}$] clouds
\item[$^{\mathrm{e}}$] adopted in the present assessment
\item[$^{\mathrm{f}}$] no other SNR with L$_{\hbox{\footnotesize 
x}}$$>$0.001 erg/s 
\end{list}

   \end{table*}

\subsection{SGR-SNR Connections}

{\bf SGR 0526-66 -- SNR N49:}
The position of SGR 0526-66 is close to the boundary of SNR N49
in the Large Magellanic Cloud. The SNR is expanding roughly spherically in a 
medium in which there are many dense clouds. Its surface brightness values
are very inhomogeneous in both radio and X-ray bands (Banas et al., 1997;
Williams et al., 1999; Castro-Tirado \& Gorosabel, 1999). As the linear
diameter of the SNR is about 15-16 pc its age should not be much
larger than t = 5.5 $\times$ 10$^3$ yrs (Vancura et al., 1992). Taking the 
age of the 
SGR to be 5.5 $\times$ 10$^3$ yrs the space velocity of the neutron star 
must be approximately 1200 km/s (Marsden et al., 1999). No bow-shock
or jet from the SGR has been observed. For comparison, PSR J1801-2451 
associated with SNR G5.4-1.2, which has the largest $\beta$ value (close 
to 1) among all of the radio pulsars connected with SNRs and a comparable
high velocity, does show a bow-shock (Frail and Kulkarni, 1991).  

N49 has a large and thermal luminosity in the ultraviolet and X-ray bands 
(Hughes et al., 1998; Banas et al., 1997). This is the result of the ambient
medium being very dense.
The explosion energy of N49 is $\sim$ 1.5 $\times$ 10$^{51}$ ergs but, as 
it is expanding in a dense medium, its expansion velocity must have 
dropped rapidly. Even with the large SGR velocity, the observed $\beta$ 
= 0.6-1 indicates that the SNR has expanded at a low rate, of the order 
of the pulsar's velocity. From the range $\beta$ = 0.6-1, V = 1000-1700 
km/s, we adopt a representative value, V$_{\hbox{\footnotesize {NS}}}$ = 
1200 km/s (Marsden et al., 1999). 
According to Allakhverdiev et al. (1997b) the space velocity 
distribution of pulsars has a mean \={V$_{\hbox{\footnotesize 
{NS}}}$} $\cong$ 250 km/s. Hansen \& Phinney (1997) give 
\={V$_{\hbox{\footnotesize {NS}}}$} $\cong$ 250-300 km/s with a standard 
deviation $\sigma$ = 190 km/s. On the other hand, Lorimer et al. (1997) 
give a higher value \={V$_{\hbox{\footnotesize {NS}}}$} $\cong$ 500 km/s. 
Although V$_{\hbox{\footnotesize {NS}}}$ = 1200 km/s for
SGR 0526-66 is $\sim$ 2.5-5 times higher than these mean pulsar velocity 
values, this velocity value is compatible with the high velocity tail of the 
space velocity distribution of pulsars, which extends to roughly 1000 
km/s. 

The N$_{\hbox{\footnotesize {HI}}}$ values of the SGR and the SNR are 
comparable within error limits and based on the positional coincidence we 
conclude that this association is real. The association will be 
definitely confirmed if future observations resolve a bow-shock or jet 
from SGR 0526-66. \\ 
{\bf SGR 1806-20 -- SNR G10.0-0.3:}
SGR 1806-20, with $\beta$ $\le$ 0.5 as seen from Table 2, is projected 
relatively close to the center of SNR G10.0-0.3. The SNR's angular 
dimensions are $6'\times9'$ (Frail et al.,
1997), so that its linear diameter must be $\sim$ 24 pc if its distance is
$\sim$ 11 kpc consistent with N$_{\hbox{\footnotesize 
{HI}}}$ = 6 $\times$ 10$^{22}$ cm$^{\hbox{\footnotesize {-2}}}$. On the 
other hand, as the X-ray surface brightness is concentrated in
the central part of the SNR (plerion type, Vasisht et al., 1995; in
Green, 2000, where SNRs are classified according to radio data only, the
type of this remnant is not determined) we can assume
that this SNR is C-type. The corresponding excess in the radio band is 
not resolved in current data.
Sometimes a jet is observed in the plerionic part of the SNR (Frail et al.,
1997). 

In Green (2000) the SNR's spectral index is $\alpha$ = 0.8 and in Vasisht et
al. (1995) it is given as $\alpha$ = 0.6. For an S- or C-type
remnant with an age of 10$^4$ yrs (Marsden et al., 1999) $\alpha$ = 0.6 is a
reasonable value. 

Even if we assume that there is an uncertainty of a factor of about 2 in the 
SNR's age the characteristic age of SGR 1806-20 is at least 3.5 
times less than the age of SNR G10.0-0.3 which corresponds to a braking 
index of 1.6. Although young radio pulsars have braking index values 
less than 3 none of them has n $<$ 2, except for the Vela pulsar for which 
n = 1.4 was claimed (Lyne et al. 1996), but this is subject to some 
uncertainty because of the large effects of interglitch relaxation in 
this pulsar.

\begin{table*}
\centering
\caption[]{The Data of Anomalous X-ray Pulsars}
\label{table3}
$$
\begin{array}{p{0.2\linewidth}ccccc}
\hline
\noalign{\smallskip}
Names \& & $P$ & $\.{P}$ 10^{-11} & \tau 10^3 & $d$ &
$N$_{\hbox{\footnotesize {HI}}} 10^{21} \\
Remarks & ($s$) & ($ss$^{-1}) & ($yr$) & ($kpc$) & ($cm$^{-2}) \\
& & & & & (0.1-2.4 $keV$) \\
            \noalign{\smallskip}
            \hline
            \noalign{\smallskip}
AXP 1E 1841-045 & 11.77 & 4.1 & 4.7 & 6-7.5 & 30 \\
SNR G27.4+0.0 & [11, 12] & [12] & & [16] & [13] \\
(Kes 73) & & & & & \\
Connection & & & & & \\ \hline
AXP AX J1845.0-0300 & 6.97 & 0.78? & 14? & \sim8.5 & 46 \\
(AX J1845-0258) & [14] & & & [14, 18] & [13] \\  
SNR G29.6+0.1 & & & & 15 & 100 \\
Connection & & & & [2] & [18, 19] \\ \hline  
AXP 1E 2259+586 & 6.98 & 0.06 & 180 & 5.6 & 8.5 \\
SNR G109.1-1.0 & [7, 5, 19] & [5, 8, 19] & & [17] & [6] \\
(CTB 109) & & & & & 9 \\
Jet [19] & & & & & [13] \\
Connection & & & & & 3-12 \\
& & & & & [20] \\ \hline
AXP RXS J170849-4009 & 11.0 & 2.25 & 8.7 & & 14 \\
SNR G346.6-0.2 & [9] & [10] & & & [13, 9] \\ \hline
AXP 1E 1048.1-5937 & 6.45 & 1.5-4 & 2.5-6.7 & >2.8, 10 & 5.5-16 \\
SNR G287.7-0.5 & [2, 1] & [3, 4] & & [2] & [1] \\
(absent in Green, 2000) & & 3.3 & 3.7 & 10.6 & 5 \\
& & [1] & & [15] & [13] \\
& & & & 3 & \\
& & & & [1] & \\ \hline   
            \noalign{\smallskip}
         \end{array}
     $$
\begin{list}{}{}
\item[$^{\mathrm{}}$]
[1] Corbet \& Mihara, 1997; [2] Seward et al., 1986; [3]
Mereghetti, 1995; [4] Oosterbroek
et al., 1998; [5] Baykal \& Swank, 1996; [6] Rho \&
Petre, 1997; [7] Fahlman \& Gregory,
1981; [8] Kaspi et al., 1999;
[9] Sugizaki et al., 1997; [10] Israel et al.,
1999; [11] Vasisht \& Gotthelf, 1997;
[12] Gotthelf et al., 1999; [13] Mereghetti, 2001a; [14]
Torii et al., 1998; [15] Mereghetti \&
Stella, 1995; [16] Sanbonmatsu \& Helfand, 1992; [17]
Hughes et al., 1984; [18] Gaensler et
al., 1999; [19] Morini et al., 1988;
[20] Parmar et al., 1998

\end{list}
   \end{table*}

\begin{table*}
\centering
\caption[]{The Data of Supernova Remnants in the Directions of AXPs}
\label{table4}
$$  
\begin{array}{p{0.1\linewidth}ccccccccccccc}
\hline
\noalign{\smallskip}
Names \& & d & t & E_{\hbox{\footnotesize {kin}}} & kT &
Type & \alpha & \Delta\theta/\theta & D &
N_{\hbox{\footnotesize {HI}}} & L_{\hbox{\footnotesize
x}} & Density & V_{\hbox{\footnotesize {NS}}} \\
Remarks & (kpc) & (kyr) & (10^{50} erg) & (keV) & & & (or \beta) &
(pc)
& (10^{21} cm^{-2}) & (10^{33} erg/s) & (\sharp/cm^3) &
(km/s) \\ \hline
            \noalign{\smallskip}
            \hline
            \noalign{\smallskip}
G27.4+0.0 & 6.5 & \le3 & & \sim0.86 & S & 0.68 & 0.1-0.2 & 7.6 &
5 & 200^{\hbox{\footnotesize b}} & Dense? & <500 \\
(Kes 73) & [5] & [5,6] & & [6] & & [16] & [5,7] & & [5] & [5] & &
[9] \\
$[7]$ & & 2 & & & & & 0.1 & & 5-20 & & & 200 \\
& & [7] & & & & & [10] & & [6] & & & [10] \\ \hline
G29.6+0.1 & 11 & <8 & & & S & 0.5? & 0.1-0.2 & 16 & & & & <500 \\
$[1]$ & 12 & [1,2] & & & & [16] & [1] & & & & & [9] \\
& [10] & 10 & & & & 0.4-0.7 & 0.1 & & & & & 200 \\
& & [10] & & & & [1] & [10] & & & & & [10] \\ \hline
G109.1-1.0 & 5 & 3 & 10-100 & 0.95 & S & 0.50 & 0.2-0.3 & 40.6
& 4 & \sim1000^{\hbox{\footnotesize c}} & D.M.C.^{\hbox{\footnotesize f}} & 
<500 \\
(CTB 109) & 3.6-5.2 & [3,4] & [14] & [4] & & [16] & [3,8] & & [14] &
\sim40^{\hbox{\footnotesize d}} & [4,16] & [9] \\
X-ray jet$^{\hbox{\footnotesize a}}$ & [15] & 10 & & 0.7 & & & 0.2 & & 
8-10 & [14] 
& \bar{n_{\hbox{\footnotesize o}}}=0.25 & \sim300^{\hbox{\footnotesize
h}} \\
$[14]$ & & [14,3,8] & & [14] & & & [10] & & [12] &
120 d_{3.5}^2{^{\hbox{\footnotesize e}}} & [14] & [10] \\
& & 12-17 & & & & & & & & & n_{\hbox{\footnotesize 
o}}=20^{\hbox{\footnotesize g}} & 15^{\hbox{\footnotesize i}} \\
& & [15] & & & & & & & & & [11] & \\ \hline
G346.6-0.2 & 10 & 20 & & & S & 0.5? & \ge1.7 & 23 & & & Dense? & 1000 \\
& 9-10 & [10] & & & & [16] & [10] & & & & Maser & [10] \\
& [10] & & & & & & & & & & [13] & \\ \hline
            \noalign{\smallskip}
         \end{array}
     $$
\begin{list}{}{}
\item[$^{\mathrm{}}$]
[1] Gaensler et al., 1999; [2] Gotthelf et al., 2000;
[3] Rho \& Petre, 1997; [4] Parmar et al., 1998; [5] Sanbonmatsu \& Helfand,
1992; [6] Helfand et al., 1994;
[7] Vasisht \& Gotthelf, 1997;
[8] Green, 1989; [9] Gaensler, 2000; [10] Marsden et
al., 1999;
[11] Fesen \& Hurford, 1995; [12] Rho \& Petre, 1993; [13] Koralesky
et al., 1998a; [14] Morini et al., 1988; [15] Gregory
\& Fahlman, 1980; [16] Green, 2000 

\end{list}
\begin{list}{}{}  
\item[$^{\mathrm{a}}$] with kT$\approx$0.3 keV
\item[$^{\mathrm{b}}$] for 0.3-4 keV
\item[$^{\mathrm{c}}$] for $>$0.1 keV
\item[$^{\mathrm{d}}$] for $>$2 keV
\item[$^{\mathrm{e}}$] for 1.2-20 keV
\item[$^{\mathrm{f}}$] dense molecular cloud
\item[$^{\mathrm{g}}$] preshock
\item[$^{\mathrm{h}}$] for t=(7-10) $\times$ 10$^3$ yrs
\item[$^{\mathrm{i}}$] for t=1.8 $\times$ 10$^5$ yrs
\end{list}

   \end{table*}

Based on the positional coincidence and the observed jet in addition to
the plerionic characteristic of the SNR, there seems to
be a genetic connection between SGR 1806-20 and SNR G10.0-0.3. \\
{\bf SGR 1900+14 -- SNR G42.8+0.6:}
SGR 1900+14, as seen from Table 2, is located outside SNR 42.8+0.6
(see also Hurley et al., 1999b).
There is a high mass X-ray binary (HMXB 1907+0.97) in this direction
(l = 43.7$^{\hbox{\footnotesize o}}$, b = 0.5$^{\hbox{\footnotesize o}}$)
with N$_{\hbox{\footnotesize {HI}}}$ $\cong$ 1.5 $\times$ 10$^{22}$
cm$^{\hbox{\footnotesize {-2}}}$ and distance $\sim$ 3 kpc (Guseinov et al.,
2000, 2001). N$_{\hbox{\footnotesize {HI}}}$ value of SGR 1900+14 is   
$\sim$ 3 $\times$ 10$^{22}$ cm$^{\hbox{\footnotesize {-2}}}$, suggesting
that it may be located in the Sagittarius arm of the Galaxy, 6 kpc from
the Sun. The distance of the SNR is given
as 5 kpc (Vasisht et al., 1994) and 6 kpc (Ankay et al., 2001)
consistent with the distance of the Sagittarius arm. \.{P} values
of the SGR vary considerably with time. An average characteristic   
age value, $\tau$ = P/2\.{P}, of 10$^3$ yrs can be adopted for this SGR. 
This value is 10 times less than the age of the SNR. As the relevance of the
characteristic age to real age depends on the theoretical model, we must
turn to other criteria before ruling out this association. If we assume the
age of the SGR to be as large as the SNR's age, i.e. 10$^4$ yrs, and if 
its   distance is 6 kpc, projection of its space velocity on the sky is
(2-3) $\times$ 10$^3$ km/s, depending on the estimation of $\beta$. These
velocities are $\sim$ 4-12 times larger than average
pulsar velocity values given above (Allakhverdiev
et al., 1997b; Hansen \& Phinney, 1997; Lorimer et al., 1997). In addition,
$\beta$ $>$ 1. Last but not least, Lorimer \& Xilouris (2000) have 
discovered a young radio pulsar, PSR J1907+0918, of period 226 ms, 
which is the likely true association of SNR G42.8+0.6. Because of these 
reasons, we conclude that there is not a 
genetic relation between SGR 1900+14 and SNR 42.8+0.6. It should also be
noted that in the direction of the SGR no SNR with
L$_{\hbox{\footnotesize x}}$ $>$
10$^{32}$ ergs/s was found in the X-ray band (Vasisht et al., 1994). \\ 
{\bf SGR 1627-41 -- SNR G337.0-0.1 (CTB 33):}
SGR 1627-41 is located outside SNR G337.0-0.1 (CTB 33) (see Table 2).   
These two objects may be approximately at the same distance. If the
SNR's progenitor was an O-type star (we may assume this to be true,
because the medium surrounding the SNR is very dense), then the SNR
is expanding within an HII region. Within the HII region there may
be some dense clouds (the SNR has an irregular shape).
The neutron star velocity values presented in Table 2 were
calculated assuming the SNR's age to be 5 $\times$ 10$^3$ yrs. As seen from
Table 2, for $\beta$ = 2-2.3 the projection of the space velocity of the
neutron star
is 800-1000 km/s. In this case, average velocity of the SNR's shock      
wave must be about 500 km/s or less. If such a young SNR's expansion
velocity were $<$ 500 km/s then the supernova explosion energy should be
relatively low.
Could the neutron star velocity be 2-4 times greater than the average
velocity of pulsars if the supernova explosion energy were really low?
This might be the case only if the supernova explosion energy was very
asymmetric.

There seems to be another contradiction concerning the SNR alone: An age
of t $\sim$ 5000 years is a large value for D = 5-6 pc. As the linear
diameter value cannot be very different from 5-6 pc, the age of the SNR
might be even less than 5000 years. In this case, the projected space 
velocity of the neutron star becomes $>$ 1000 km/s, making the 
discrepancy between the neutron star kinetic energy and the supernova 
explosion energy even stronger.
In addition to these difficulties, $\beta$ = 1.6-2.3, i.e. the neutron
star is well outside the remnant. A genetic connection between SGR
1627-41 and SNR G337.0-0.1 is unlikely.

\subsection{AXP-SNR Connections}

{\bf AXP 1E 1841-045 -- SNR G27.4+0.0:}
AXP 1E 1841-045 is projected on the central region of SNR G27.4+0.0
(see Table 4). It is seen from Tables 3 and 4 that the 
N$_{\hbox{\footnotesize {HI}}}$ value of 
the AXP is about 3 times greater than the SNR's neutral hydrogen column
density. If we take into account the uncertainty in 
N$_{\hbox{\footnotesize {HI}}}$ values we cannot
exclude the possibility of a connection between the AXP and the SNR. If 
the AXP is a magnetar, the characteristic age of the AXP is 4.7 
$\times$ 10$^3$ yrs. The age of the SNR is estimated as (2-3) $\times$ 
10$^3$ yrs. Considering the uncertainties in the ages of the AXP and the 
SNR we can assume that both have ages of about (3-5) $\times$ 10$^3$ 
yrs. This is not a bad assumption, because 
the implied space velocity of the AXP is $<$ 500 km/s (Table 4, 
last column). So, AXP 1E 1841-045 might be born in the same supernova 
explosion that formed the SNR G27.4+0.0 which is considered to be
an S-type SNR (Green, 2000). 

On the basis of the central position of AXP
1E 1841-045 within SNR G27.4+0.0 and the compatibility of 
distance values, we conclude this is a true association. The ages are 
also consistent if the AXP is a magnetar. \\
{\bf AXP J1845.0-0300 -- SNR G29.6+0.1:}
AXP J1845.0-0300 is projected on the center of SNR G29.6+0.1.
This AXP has an N$_{\hbox{\footnotesize {HI}}}$ value of $\sim$ 
(5-10) $\times$ 10$^{22}$ cm$^{\hbox{\footnotesize {-2}}}$. SNR G29.6+0.1 
is at $\sim$ 11 kpc and has a direction of l $\sim$ 30$^{\hbox{\footnotesize 
o}}$, so that, the line of sight
cuts the Sagitarius arm of the Galaxy twice and the Expanding arm,
which is 3 kpc away from the Galactic center, once. So, 
the N$_{\hbox{\footnotesize {HI}}}$ value of 
the SNR can also be large, possibly of the order of the AXP's 
N$_{\hbox{\footnotesize {HI}}}$. For such high N$_{\hbox{\footnotesize 
{HI}}}$ values it 
is not possible to test distance agreement to better than 2-3 kpc.
Thus, we can not exclude the possibility of a connection between the AXP
and the SNR, because the AXP is projected on the center of SNR G29.6+0.1.
SNR G29.6+0.1 has not yet been investigated in detail. There may be
structure in the central region of this remnant. If resolved this would 
constitute supportive evidence for the association. We conclude that a 
genetic connection between AXP J1845.0-0300 and SNR G29.6+0.1 is likely. \\ 
{\bf AXP 1E 2259+586 -- SNR G109.1-1.0 (CTB 109):}
AXP 1E 2259+586, as seen from Tables 3 and 4, must have a genetic   
relation with SNR G109.1-1.0 (CTB 109) (see $\beta$, d, 
N$_{\hbox{\footnotesize {HI}}}$ values). This
SNR is S-type in both radio and X-ray bands. In this region of the Galaxy,
in the anticenter direction, it is particularly improbable for an AXP to 
have a projection on the central region of a SNR by chance. As evidence for 
connection, X-ray observations show a jet in CTB 109 (Morini et al., 1988). 

Contrary to these indications of a connection, if the AXP is a magnetar the 
SNR's age of 
about (3-10) $\times$ 10$^3$ yrs does not agree with the AXP's characteristic
time P/2\.{P} = 1.8 $\times$ 10$^5$ yrs (for n = 3).
As known from different evolutionary tracks of pulsars on the P-\.{P}     
diagram the braking index (n) of a pulsar which is expected to have a very 
high magnetic field and has a luminosity of about 10$^{35}$ ergs/s in the
X-ray band may be $\le$ 3 because of the interaction of the pulsar with the
surrounding matter (Yusifov et al. 1995). But, in this case the difference 
between the ages
becomes even larger. Even if we assume that the AXP is a magnetar with 
significant field decay, and the SNR's age has an uncertainty of a factor 
of $\sim$ 2, an average braking index n $\ge$ 13 is obtained. For such 
power law decay, or exponential decays of the magnetic field, we might 
expect to see evidence of the energy released, since magnetar fields are  
supposed to be responsible for the bursts in SGRs. There is no evidence 
of an extra energy source in the properties of AXP 1E 2259+586 and SNR 
CTB 109. 

It is highly probable that there is a genetic relation between
AXP 1E 2259+586 and SNR G109.1-1.0 on the basis of the significant 
positional coincidence and distance. In this case, there is a strong age 
discrepancy for the magnetar model. \\ 
{\bf AXP RXS J170849-4009 -- SNR G346.6-0.2:}
For this AXP -- SNR pair, the value of $\beta$ $\ge$ 1.7 is large.
If the distance of the AXP is 10 kpc and its age is 2 $\times$ 10$^4$ 
yrs, the same as the SNR's age, projection of its space velocity
on the sky will be 1000 km/s. If the angular diameter of the SNR is $8'$
(Green 2000) then, from the $\Sigma$-D relation, its diameter and
distance will be 23 pc and 10 kpc, respectively (Ankay, 2001). At 10 kpc  
the AXP would be beyond the Inner (Expanding) arm of the Galaxy. The 
N$_{\hbox{\footnotesize {HI}}}$ value is consistent with such a location.
Although distance and N$_{\hbox{\footnotesize {HI}}}$ values are 
compatible, $\beta$ is so large, and the velocity inferred is also large 
compared to the velocities inferred for the other AXPs. We conclude that 
there is no connection between the AXP and the SNR. \\
{\bf AXP 1E 1048.1-5937 -- G287.8-0.5:}
There is no candidate SNR that can be genetically related with AXP 1E
1048.1-5937. In the direction of this AXP there is the source G287.8-0.5  
which was given as an unexamined SNR candidate (Jones,
1973; Becker et al., 1976), but it is not included in the Galactic
SNRs catalog of Green (2000). As the data of this source are insufficient
and not reliable it is difficult to claim it as a SNR and it does not make   
much sense to examine its possible relation with the AXP. \\
{\bf AXP 4U 0142+61:}
There is no SNR observed in this direction.

\section{Conclusions}

We found that 3 AXPs and 2 SGRs are, most probably, genetically connected
with related SNRs: AXP 1E 1841-045 -- SNR G27.4+0.0 (Kes 73), AXP AXJ 
1845.0-0300 --
SNR G29.6+0.1, AXP 1E 2259+586 -- SNR G109.1-1.0 (CTB 109), SGR 0526-66 -- 
SNR N49, SGR 1806-20 -- SNR G10.0-0.3.

The difference between the age of the SNR and the characteristic time of
the AXP/SGR is very large in two cases even if we take into account the 
uncertainty in the age of the SNR which is about a factor of 2. For the pair 
AXP 1E 2259+586 -- SNR
G109.1-1.0 (CTB 109) the $\tau$ value of the AXP is at least 6 times greater
than the age of the SNR. So, the braking index of AXP 1E 2259+586 must be at 
least 13. In the case of SGR 1806-20 -- SNR G10.0-0.3 the
situation is just the opposite; the age of the SNR is at least 3.5 times
greater than the $\tau$ value of the SGR which corresponds to a 
braking index value of about 1.6. For the other two associated 
AXP-SNR pairs, the $\tau$ and SNR age values are comparable within error 
limits. These characteristic ages are relevant to magnetar models. The 
associations then indicate that, if the AXPs and SGRs are magnetars then not 
all AXPs and SGRs are the same kind of magnetar. Furthermore, if the supposed 
magnetar age is as small as the SNR age, as in the case of AXP 1E 
2259+586, then the rate of energy release by the magnetic field is very 
large. The question is then why there is no evidence of this energy 
release either in the SNR or in some signature of the AXP.

SGR 0526-66 and SGR 1806-20, which are associated with SNRs, have 
projected space velocities V$_{\hbox{\footnotesize {NS}}}$ $\sim$ 800-1200 
km/s, higher than the average space velocity of pulsars but consistent 
with the high velocity tail of the radio pulsar space velocity distribution. 
The AXPs connected with SNRs have projected space velocities $<$ 500 km/s  
consistent with the average space velocity of pulsars. We note that the
velocity values given in Tables 2 and 4 refer to the projected space 
velocity on the sky and, on the average, the space velocity is equal to 
(3/2)$^{1/2}$ times the projected space velocity.

The SNRs connected with the SGRs are not normal S-type remnants, whereas
the SNRs connected with the AXPs are pure S-type. Similar to the 
effects of radio pulsars connected with SNRs, the effects of SGRs on
their related SNRs are stronger compared to the effects of
AXPs. For pulsars the source of such
effects on SNRs is the loss of
the neutron star's rotational energy,
i.e. $\dot{E}$ = 4$\pi$$^2$I($\dot{P}$/$P$$^3$). For the effect
on the SNR to be large $\dot{E}$ must be large. 
$\dot{E}$ values of AXPs and SGRs are about
10$^2$-10$^3$
times less than the $\dot{E}$ values of young pulsars. Because of
this, the SNRs genetically related with AXPs and SGRs being S-type would 
be considered normal if rotational energy were the only energy source. If 
AXPs and SGRs are the same type of object, and if they have energy 
sources other than rotation, either for accretion or for magnetar models, 
why is it that the SNRs associated with SGRs are different from the SNRs 
associated with AXPs? In particular, if AXPs also 
experience gamma-ray bursts powered by magnetic energy, then why is it 
that the SNRs associated with AXPs do not contain plerions?Are the 
differences in the SNR morphologies significant, since we have only five 
SNRs associated with AXPs and SGRs? These issues may be resolved with 
better resolution images of the SNRs.

We investigated the medium in which the AXPs, SGRs, and
pulsars (genetically related with the SNRs) are located in to find possible
masses of the progenitors through associated star clusters and to check if 
there are some differences
in density of the ambient medium. Contrary to the claim of Marsden
et al. (2001), we did not find any significant differences in the density
of the SNRs connected with AXPs/SGRs compared to the density of the
SNRs associated with radio pulsars located in Galactic arm regions (see
Tables 2 and 4).                                                              

\begin{acknowledgements}
We thank T\"{U}B\.{I}TAK, the Scientific and Technical Research Council of 
Turkey, for support through TBAG-\c{C}G4 and through the BDP program for 
doctoral research. AA thanks T\"{U}B\.{I}TAK for graduate student 
scholarship. \end{acknowledgements}

\end{document}